\documentclass[]{elsart}

 \usepackage{umlaut}
 \usepackage{graphicx}

\usepackage{amssymb}

\begin{document}

\begin{frontmatter}

\title{Racoon: A Parallel Mesh-Adaptive Framework for Hyperbolic Conservation Laws}

\author{J. Dreher}
\ead{dreher@tp1.rub.de}
and
\author{R. Grauer}
\ead{grauer@tp1.rub.de}

\address{Theoretische Physik I, Ruhr-Universit{\"a}t Bochum, D-44780 Bochum, Germany}

\begin{abstract}

We report on the development of a computational framework for the
parallel, mesh-adaptive solution of systems of hyperbolic conservation
laws like the time-dependent Euler equations in compressible gas
dynamics or Magneto-Hydrodynamics (MHD) and similar models in plasma
physics.
Local mesh refinement is realized by the recursive bisection of grid
blocks along each spatial dimension, implemented numerical schemes
include standard finite-differences as well as shock-capturing central
schemes, both in connection with Runge-Kutta type integrators.
Parallel execution is achieved through a configurable hybrid of
POSIX-multi-threading and MPI-distribution with dynamic load
balancing.
One- two- and three-dimensional test computations for the Euler
equations have been carried out and show good parallel scaling
behavior.
The {\em Racoon} framework is currently used to study the formation of
singularities in plasmas and fluids.

\end{abstract}

\begin{keyword}
AMR \sep mesh refinement \sep hybrid parallelization \sep multithread \sep MPI  \sep load balancing
\end{keyword}

\end{frontmatter}

\newcommand \Rac {{\em Racoon~}}

\section{Introduction}

Adaptive mesh refinement (AMR) techniques have become increasingly
popular in recent years for the numerical investigation of dynamic
phenomena in fluid dynamics and plasma physics.
The underlying idea, as described by Berger and Colella in the context
of shock hydrodynamics \cite{BerCol89}, is to provide a certain
spatial resolution of the computational grid that is necessary for the
desired accuracy of the solution {\em locally} by introducing a
hierarchy of subgrids of different sizes and spatial resolutions to
cover different regions of the domain.
As many problems involve or develop {\em localized} small scale
structures, it is then possible to compute such critical regions on
finer grid blocks while other parts of the domain are well represented
on coarser blocks, which can drastically reduce the computational
costs compared to using a fine grid in the entire domain.
As the system evolves dynamically, with small-scale structures likely
to form, move and disappear, their coverage by the recursively refined
blocks must be adjusted appropriately in order to ensure proper local
resolution.
Following this strategy, the computational expenses in terms of memory
and CPU usage can easily be reduced to a few percent compared to
corresponding simulations based on fixed grids in which the resolution
is dictated by the finest structures \cite{Kep2002} whenever the
phenomena under investigation exhibits strongly localized structures
like shocks, near-singularities, boundary- or interface layers and the
like.

Successful fluid dynamical and plasma physical applications of this
approach today already cover a wide spectrum, including the problem of
magneto-convection in the solar atmosphere \cite{Steiner}, the
formation of singular current sheets in magneto-hydrodynamics
\cite{Friedel},\cite{GrauMar} and vortex tubes in hydrodynamics
\cite{GrauMarGer}, the propagation of solar coronal mass ejections
through the heliosphere and their interaction with the Earth's
magnetosphere and ionosphere \cite{Groth}, and accretion phenomena in
astrophysics \cite{Ziegler2001}.
As AMR is merely a supportive technique to improve the economy of
computations, a more widespread usage in many different fields can be
expected for the future, which, of course, is connected to the
progress being made in related areas, e.g. the use of refined physical
models, the improvement in numerical methods, and the gain in
computational power through modern parallel hardware architectures.

Given these promising perspectives on the one hand and the fact that
significant progress occurs in the quite specialized disciplines named
above, an obvious challenge is the development of programming
environments that combine such modern computational techniques with a
degree of flexibility that is necessary to avoid highly specialized,
and thus expensive, monolithic solutions in favor of easy adoption,
extension and maintenance.
Recent steps into this direction are the development of a number of
libraries that are specifically targeted at the grid handling tasks in
AMR computations like the Fortran-based PARAMESH
\cite{MacNeice}, the freely available DAGH library, written in C++ 
\cite{DAGH}, and the object-oriented designed  SAMRAI framework \cite{SAMRAI}.
All of these support parallel execution in distributed memory
architectures.
Other codes have been developed with more or less specific physical
applications in mind, for instance the FLASH code for astrophysical
phenomena \cite{Flash}, which is built on top of the PARAMESH library,
and the Versatile Advection Code by T{\'oth} \cite{Toth},
which started as a collection of different numerical schemes for fluid
and magneto-fluid dynamics, formulated in a dimension-independent
fashion, and was recently extended to offer AMR capabilities
\cite{Kep2002}.
This code uses OpenMP parallelization and is thereby restricted to
(virtually) shared memory systems.
The same applies to the incompressible fluid dynamics code used by
Grauer et al. for the study of the formation of singular structures
\cite{GrauMar}, \cite{GrauMarGer}, in which multi-threading is
implemented explicitly through the POSIX standard interface.

In the present paper, we describe \Rac, a framework that has
recently been developed to offer an environment for the grid-adaptive
solution of conservative systems and related systems.
The major motivations here were the possibility to efficiently exploit
both shared and distributed memory architectures, and to keep the
design as flexible as possible with respect to the numerical schemes
used, the physical problems addressed, and forthcoming extensions for
more sophisticated simulation techniques.
\Rac is implemented in C++ and offers the necessary
functionality for computations in 1 to 3 spatial dimensions with the
possibility for extensions into higher dimensions.

In the next section, we start with a few general remarks on the
overall design philosophy and a brief description of the numerical
schemes that have been implemented by now.
The grid refinement algorithm and the resulting data communication
patterns during the computation are outlined in section~3.
Section 4 is devoted to the parallelization strategy, which is a
hybrid of multi-threading and interprocess communication through the
Message-Passing Interface (MPI), and to some benchmarking results of
the scaling in parallel environments.
Finally, the conclusion summarizes the key findings and gives a critical 
assessment of the mixed-mode parallelization.

\section{Design Aspects}
\label{design}

Obviously, an AMR framework, in particular when designed for
distributed parallel architectures, exhibits considerably more
complexity than conventional computations on fixed grids.
Therefore, proper modular code design seems even more advisable than
in traditional computations, and the use of corresponding techniques
from software engineering that have emerged over the last decades or
so, is at least worth being considered.
While the adoption of object-oriented design concepts in high
performance computing has been somewhat reluctant over the years,
there now seems to be a clear tendency to provide at least C++
interfaces to the many standard libraries in this field or even to
provide the implementation directly in C++.
Reasons for this tendency might be that the run time penalty of C/C++
codes compared to Fortran codes continues to diminish and, in some
cases, has entirely disappeared, in particular when using restricted
pointers for de-aliased data access.
In fact, some of the AMR frameworks mentioned in the introduction
already exploit the potential of object-oriented formulations
\cite{SAMRAI}, \cite{DAGH} for the operations on grid data and for the
data communication between individual grid blocks.

\Rac has been implemented entirely in C++, since this language
offers various mechanisms which allow a fine-grained control over the
code generation:
It supports full runtime polymorphic behavior, but offers also
``non-virtual'' object-based programming with full compile time
resolution which allows function inlining and similar optimizations.
In addition, template programming is a technique that combines high
abstraction without necessarily sacrificing much runtime performance.
Widely used numerical libraries for e.g.~linear algebra or FFT are not
applicable in the present context as the solution of hyperbolic
systems on locally refined meshes with explicit time stepping is a
local problem not covered by those algorithms.

In contrast, the numerical schemes employed here, are highly tailored
towards conservation laws like compressible gas dynamics or
Magneto-Hydrodynamics.
They operate on regular numerical meshes and basically involve the
computation of numerical fluxes for physical quantities at cell
boundaries and an update of the cell integral according to the time
integration scheme.

At present, \Rac implements the Kurganov-Tadmor \cite{KT2000} scheme, which is a
successor to the one originally proposed by Nessyahu and Tadmor
\cite{NT90}, and falls into the class of Lax-Friedrichs type schemes
\cite{LxF}.
A major feature here is that they are much simpler to
implement than classical Godunov-type schemes \cite{Godu},
\cite{WoodPPM}, \cite{WAF}, as they avoid the spectral decomposition
of the conservation equations, and hence are applicable in situations
where (approximate) Riemann solvers \cite{Roe} are unavailable or too
expensive.
The Kurganov-Tadmor version in particular operates on a non-staggered mesh,
making it well-suited for AMR computations.
Used in connection with a CWENO-reconstruction, and an appropriate
Runge-Kutta integrator, it offers third order accuracy in smooth regions
in both space and time.
As a cheaper and less powerful alternative, standard second- and fourth-order
finite difference discretizations are also implemented.

The entire framework consists of the following major functional components:
\begin{itemize}

\item
Data management on the grid blocks: Storage, allocation, inter-block communication
(e.g., data copy, interpolation), geometry information, block creation
and disposal.

\item
Numerical treatment of a single grid block by means of a given scheme
and problem (time stepping, computation of numerical fluxes)

\item
Definition of the physical system: Variables, flux function (possibly
source terms), initial values, boundary values on the physical domain,
diagnostics etc.

\item
Mesh refinement, regridding, load balancing and -distribution.

\item
Parallelization: Interprocess communication (MPI), thread synchronization etc.

\item
Overall control flow.

\end{itemize}

Data for physical quantities are kept in standard C-arrays on each
individual grid block, and all numerical operations are executed on
these arrays with the help of meta-information like array dimensions,
number of overlapping ghost points and the like.
Comparisons with available libraries like Blitz++ has shown that they
offer no runtime gain compared to the solution chosen here.
Every block has a unique identity that specifies the blocks position
and size in the domain and keeps direct references (pointers) to its
logical neighborhood, i.e.~neighbor blocks, the overlayed parent
block and its child blocks, as far as they exists.
The blocks in turn are organized in a hierarchical manner within a
``Grid'' object, which basically consists of lookup-tables for the
blocks in each refinement level and cached block connectivity
information for faster grid traversal.

Numerical schemes are predefined for 1- to 3-dimensional computations
and act on a configurable subset of the physical fields.
The calculation of the fluxes and sources, boundary values at the
physical boundaries etc.~occurs through callback functions of
appropriate C++ interfaces. For this purpose, a specialized
``Problem'' instance, that either directly offers the necessary
information about the model problem or delegates to other objects,
registers itself at program start-up and gets called by the main
process control, schemes etc.~for the necessary action.
Details of the mesh refinement algorithm and the parallelization are
given in the following sections.

\section{Mesh refinement}

\subsection{Mesh Refinement Strategy}
In their pioneering paper, Berger and Colella \cite{BerCol89} describe
a patch-based mesh refinement method for hydrodynamic computations
with shocks or highly localized structures:
Starting the simulation with a single grid of a given initial
resolution, an a-posteriori error estimate is used to identify
critical points on the mesh with insufficient spatial resolution.
Then, rectangular grid patches of smaller grid spacing and
correspondingly higher resolution are introduced to cover those
critical points and guarantee the desired accuracy.
By applying this method recursively, a hierarchy of grid patches is
created so that areas featuring small scale structures are resolved as
needed, while the solution is represented and computed on much coarser
grids in smooth regions.
We call this approach patch-based because the shapes, sizes and
positions of the subgrids don't follow simple rules and depend on the
temporal evolution of the system under investigation: Constraints are
only their rectangular shape and some nesting conditions which
basically state that every grid patch, supplemented with a border
region, must lie entirely within an area already covered by patches of
the next coarser level.
The attractiveness of this approach is the efficient coverage of small
scale structures: Once critical points are identified, the grid
patches are computed to give an optimal coverage of critical regions.

While this method has been successfully used and implemented in a
number of later codes \cite{Friedel}, \cite{Steiner},
\cite{Kep2002}, there are some drawbacks:
The algorithm to compute the shape and position of the patches under
the nesting constraint is far from trivial.
Moreover, the fact that the subgrids created this way are irregular in
size, shape and position and may abut each other in unpredictable
ways means that the regridding procedure for a given refinement level
$l$ affects all patches on this level as a whole, making it a
non-local task.
This, together with the equally irregular data exchange between the
subgrids, poses severe difficulties for the parallel execution in
distributed memory architectures.

An alternative to this patch-based approach is the regular bisection
of grid blocks along each spatial dimension, as used in more recent
AMR implementations \cite{MacNeice}, \cite{Powell}:
Instead of collecting critical points into irregular patches, every
grid block that contains at least one critical point is refined as a
whole by shadowing or replacing it with a number of equally sized
squares (2D) or cubes (3D), which are created by bisecting their
``parent'' block along every coordinate direction.
Using this refinement ratio of 2, a $d$-dimensional implementation
replaces a given block of level $l$ by its $2^d$ children of level
$l+1$.
If the number of cells per block and direction is the same on all grid
blocks, the grid resolution increases by a factor of 2 with increasing
level $l$.
An illustration of this ``regular'' refinement in 2D is given in
Figure 1, and more advanced methods, for example with the option of
anisotropic refinement along only one direction or other refinement
ratios than 2, can be found in the literature.

\begin{figure}
\centerline{\includegraphics[width=9cm]{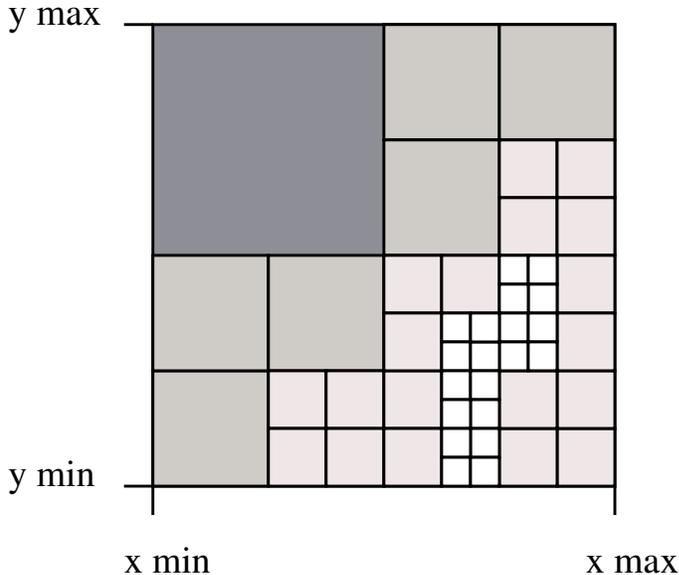}}
\caption
{
Illustration of regular grid refinement: The
domain is covered by a hierarchy of grid blocks, each of them
hosting a fixed number of cells.
Upon refinement, a block is shadowed by 4 (in 2D) or 8 (in 3D)
equally sized children.
Shown here are refinement levels between 1 (upper left) and 4.
}
\end{figure}

As a consequence of this rather simple approach, a logical
tree-hierarchy of grid blocks is created that, despite a less
efficient coverage of critical points, has a number of favorable
advantages:
Well-defined relationships between blocks exist in terms of
parent-child relations between refinement levels and neighbor
relations on the same level, which makes the implementation of data
transfer operations like data copy, interpolation and averaging,
relatively easy.
Common boundaries shared by neighbor blocks coincide entirely, and a
parent's boundaries are contained in the union of its children's
boundaries.
Due to the tree hierarchy, ordering and traversal of the blocks is
straight-forward and block sizes and positions within the domain are
well-defined.

This regular subgridding approach was adopted in \Rac because of its
compelling simplicity of the regridding procedure, the easy
formulation of data communication operations and, in particular, the
advantages that the logical block order offers for the parallel
implementation, including the load balancing.
In fact, we use a compound block identifier $g=(l, i)$ to assign to
every grid block a globally unique identity consisting of the
refinement level $l$ and a running index $0 \le i < 2^{l*d}$.
With $g$ alone, a block's spatial position and size are determined, as
are the corresponding ID's of its neighbors, parent and children if
they exists.
The mapping between the index $i$ within a given level $l$ and the
location of the grid block $(l, i)$ in the domain, i.e. the numbering
of blocks, is induced by space-filling curves of the Hilbert type
\cite{Hilbert}, a choice which is made in order to obtain an efficient
load distribution strategy for distributed parallel processing as
described in section 4.

An apparent penalty compared to the patch-based method is the fact
that the coverage of critical sections at first sight seems less
efficient with grid block bisecting because the border lines of
possible refinement level boundaries are defined a priory.
However, this turns out to be less severe than expected if the
computations are started with a refinement level $l>0$ right away,
i.e. with a domain decomposition: Then, only few further refinements
are needed to localize the high resolution regions around critical
structures.
In addition, a certain margin of increased resolution around critical
structures is necessary to accommodate their possible motion relative
to the grid between regridding phases.
Typical runs start with $l=3$ or $l=4$, corresponding to a domain
decomposition into 64 or 256 blocks in 2D and 512 or 4096 in 3D.

An important positive side effect of starting the simulation with a
domain-decomposed setup right from the beginning is the gain in
execution speed:
Even in sequential mode, the decomposed domain with relatively small
grid blocks (typically $8^d$ to $16^d$ cells) leads to a much better
CPU cache utilization than one large grid block, and this cache
efficiency easily comes with a gain of factor 2 or more in execution
speed on the tested platforms (i686, Opteron, power4) despite the 
necessary boundary data copy.
In parallel mode, the coarsest grid level is easily parallelized.

A regridding procedure is initiated after a configurable number of
time steps by checking every point in every block for a given
refinement criterion.
The criterion itself is defined together with the physical problem and
is typically based on the gradient of some variables as to identify
small scale structures in the domain.
A more sophisticated and possibly less biased approach might be a
thorough error estimate, for instance using Richardson-extrapolation
as proposed by Berger and Colella \cite{BerCol89}.
Grid cells which are found to be critical cause the containing blocks
to be flagged for refinement.
In addition, those critical points which are close to the block's
boundary are used to flag the abutting neighbor block(s) as well in
order to communicate the information about critical structures across
block boundaries and ensure a buffer zone of high resolution around 
critical points.
Finally, additional refinement marks are set in order to ensure a grid
consistency criterion which states that abutting blocks may differ by
at most one refinement level. 
This criterion greatly simplifies the interpolation operation at
refinement boundaries and comes at almost no cost.

After this marking step, the entire grid is updated:
Flagged blocks, unless already refined, are supplemented by their
children, using the actual parent block's data for initialization,
while unnecessary refinements are removed by deleting the
corresponding child blocks after their actual data was transfered
to the coarser parent block.
The parent blocks, even if refined, are still kept in the computation.
Depending on the integration and time stepping mode that was selected,
they are either used to obtain a first approximation during every
integration step and provide intermediate boundary values for children
and neighbors, or they are simply deactivated and kept passively in the
process.
Overall bookkeeping of the blocks is handled by ``grid'' objects, as
already mentioned, where the block ID's (level-key pair) are used as
keys and iterators are provided for the traversal of the entire grid
or blocks of a certain refinement level.

\subsection{Time stepping}

Blocks are integrated quasi-autonomously, which requires bands of
ghost cells around the actual physical area that have to be updated
after every integration step:
For neighboring blocks of the same refinement level, this is just a
copy of the overlap region like in a domain decomposition, and blocks
which abut the physical domain boundary are treated according to the
corresponding physical boundary condition.
At refinement boundaries, finer blocks receive interpolated ghost cell
values from their respective coarser neighbors, while coarser blocks
receive averaged values from the fine region.

In order to fulfill the Courant-Friedrichs-Levy (CFL) condition for
numerical stability, finer grid blocks will in general need a smaller
integration time step than their coarser counterparts.
In most applications, this stability restriction is more relevant than
the accuracy condition due to the explicit solvers used.

There are two basic ways for the time stepping on locally refined
grids:
1. One common step is used for all grid blocks, dictated by the
most stringent conditions found in the entire domain, or 
2.  different refinement levels use individual time steps so that
coarser blocks in general will be integrated with fewer larger steps
than fine blocks, which leads to a {\it temporal refinement} in
addition to the spatial one.
The main reason for using the first method is it's simplicity and, for
more advanced systems like incompressible flows, the compatibility with
additional correction steps (e.g.~ velocity projection after solving a
pressure equation).
The advantage of the second approach is, apart from moderate savings
in computing time, the fact that also on coarse blocks, the CFL number
can stay close to 1 which is favorable for the phase error confinement
in some numerical schemes.

One of either methods can be selected in the \Rac time stepping module,
and the time step condition itself is always provided by the specific
problem instance which gets periodically queried by the time step
classes which in turn calculate the individual time steps for each
refinement level and the resulting integer factors between them.

When individual time steps $\Delta t_l$ are chosen for each level $l$,
one step on the entire grid is formulated as a level-wise, recursive
procedure (intra-level communication with data copy is left out for
clarity):
\begin{itemize}
\item
{\it step grid blocks on level $l$, individual $\Delta t_l$:}  
\begin{itemize}
\item
save data at refinement boundaries to level $l+1$ (old)
\item
integrate level $l$ blocks once with time step $\Delta t_l$
\item if not coarsest level: update boundary values at boundary to
coarser blocks as temporal interpolation
\item pass spatial interpolation of old an new data at boundaries to
level $l+1$ to corresponding level-$l+1$ blocks
\item
call {\it step grid blocks of level $l+1$} $\Delta t_l/\Delta t_{l+1}$ times
\item
if last step on level $l$, pass latest coarsened volume data to level
$l-1$-blocks at refinement boundaries
\end{itemize}
\end{itemize}
This methods requires that shadowed coarse grid blocks at boundaries
to finer blocks are included in the integration in order to provide
temporal interpolations of boundary data for the abutting finer blocks.
It is evident that the algorithm results in a level-by-level
integration with intermediate data transfer between levels.

If one common time step is selected, the 
procedure is considerably simpler:
\begin{itemize}
\item
{\it step all grid blocks with common $\Delta t$:}  
\begin{itemize}
\item
step all blocks once with time step $\Delta t$
\item
pass coarsened boundary data from fine to coarse levels at refinement
boundaries, starting with finest level
\item
pass interpolated boundary data from coarse to fine levels at
refinement boundaries, starting with coarsest level
\end{itemize}
\end{itemize}
Even here, the boundary data exchange has to occur separate for the
fine-to-coarse direction and vice-versa due to the implicit data
dependency:
For the calculation of the (new) local interpolation to be passed to
the fine ghost cells, the coarse blocks must have new values in their
own boundary (ghost) cells already available.

While it would be possible in principle, to keep track of the data
flow for each individual block and to use the data dependency as a
criterion to start integrating a block, we have followed the approach
of a level-by-level global synchronization.
This facilitates the entire flow control and seems by far the most
practical way to go as the dependency graphs between blocks get
arbitrarily entangled after local refinement (note that data is
exchanged not only between face neighbors, but, depending on the
numerical scheme, also diagonal neighbors; interpolation depends on
previous intra-level exchange etc.).
In particular in the case of multi-threaded execution to be described
in the next section, this data tracking would require additional
thread-synchronization for consistent housekeeping, and we expect that
the overall gain would be small compared to the necessary effort.
The global synchronization doesn't require a separate global
communication action: It's simply provided by synchronizing threads
within each process after all external communication has been finished.

\section{Parallelization}

High performance computing today requires efficient parallel
execution.
Adaptive mesh refinement suggests the fundamental parallelization
strategy of handling every single grid block as a quasi independent
piece of work for each integration step, and assign the blocks to CPUs
for integration.
The existence of ghost cells around the blocks, which effectively
represent a buffer zone, allows this approach to be implemented
directly.
However, it must be kept in mind that AMR execution, and thereby its
parallelization, is considerably more complex than a pure domain
decomposition:
Blocks are recursively advanced level by level as described in chapter
3, and information like interpolated boundary values and flux
corrections are exchanged between parents and children at level
changes.
In shared memory architectures, the corresponding communication paths
are still workable without particular consideration: For instance,
Keppens and T{\'o}th \cite{KepOMP} report run time speed-up factors of
up to 4.9 on a 16 CPU SGI Origin using manual parallelization with
OpenMP.  These results are in agreement with the scaling that Grauer
et al. \cite{GrauMarGer} obtained using multi-threading.

However, as the trend in high performance computing moves towards
cluster architectures, distributed parallelism must be addressed in
order to exploit these platforms.
The approach of regular subgridding as presented in chapter 3 is mainly
motivated to ease distributed parallelism:
An obvious advantage here is that the regular interfaces between
neighboring (and nested) blocks facilitate the programming, making a
distributed implementation through the Message Passing Interface (MPI)
workable.
Moreover, equal block sizes mean roughly the same computational load
per block, which greatly facilitates the dynamical load balancing.
A further aspect is the fact that we can globally identify a grid
block by means of its block ID, consisting of the refinement level and
a running index, across address spaces, which makes it easy to send
block-to-block messages between computing nodes.
Finally, the indexing of blocks of a given level by means of Hilbert
curves, as already mentioned in chapter 3, offers a distribution
strategy that aims at minimizing the inter-node communication.

With \Rac, an attempt has been made to combine multi-threading with
MPI distribution in order to adopt to cluster architectures consisting
of SMP nodes.
In the following subsections, this approach will be described in
detail together with benchmark results of the parallel performance.

\subsection{Hilbert Curve Ordering of Blocks and Load Distribution}

As already mentioned, Hilbert type space filling curves are used to
order and distribute the individual grid blocks in a way which aims
at minimizing the communication between concurrent threads of program
execution and between processes.
Hilbert space filling curves of level $l$ in $d$-dimensional space
provide a mapping between the interval $[0,1]$ and the domain
$[0,1]^d$ with the property that neighborhood or proximity tends to be
conserved under this mapping.
This means that points being close to each other in the interval
$[0,1]$ tend to be mapped to points in space which are also have a
small distance in $R^d$ and vice versa, a feature which is utilized in
several areas of computer science.
For illustration, the 2D Hilbert curves of level 1 and level 2 are shown 
in Figure 2.
\begin{figure}
\includegraphics[width=6.5cm]{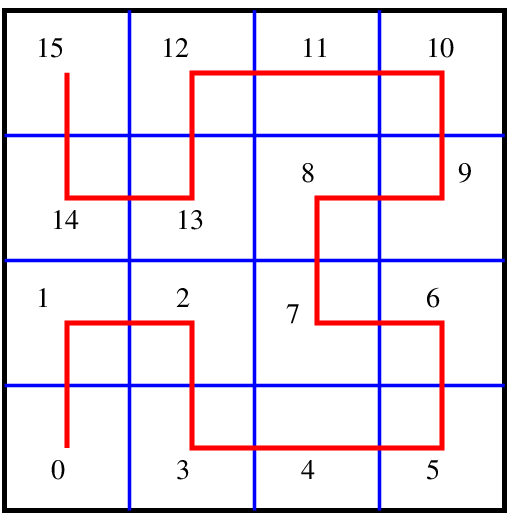} \hspace*{0.5cm}
\includegraphics[width=6.5cm]{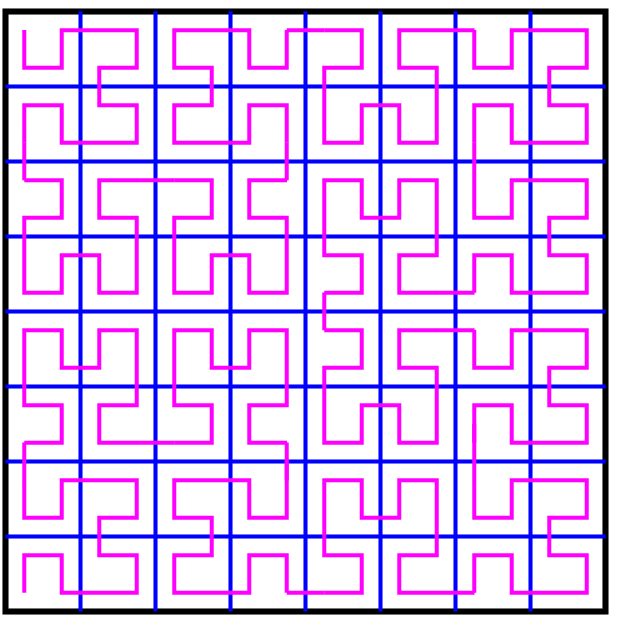}
\caption
{ Two-dimensional Hilbert curves of level 2 (left) and 3 (right),
respectively. Each of the squares corresponds to a potential
grid block. Blocks of a given level are ordered according to their
position on the visiting Hilbert curve in order to achieve clustering
on computing nodes.  }
\end{figure}
This tendency of neighborhood conservation is used to determine the
distribution of blocks among separate computing nodes
(processes/threads) in a way that tends to assign neighboring blocks
to the same node and thereby minimize inter node communication.
A similar approach is used in the PARAMESH library by MacNeice et
al. \cite{MacNeice} and theoretical investigations into the efficiency
of Hilbert curves in load distribution are given in e.g.~\cite{Zumbusch2001}
and \cite{Zumbusch2003}.

To achieve this communication-efficient distribution, blocks of a given
level $l$ are ordered according to their Hilbert index $i$, and this
sequence is divided into equally sized partitions (up to rounding
error) for computing nodes.
With this mechanism, every block is assigned to a node, and
physically close regions of the domain tend to end up on the same node.
It should be noted that it is also possible to formulate a order not
only per level but of all blocks, which additionally conserves the
proximity between parents and children.
This, however turned out to be disadvantageous for our application:
While computational load balance can be achieved by weighting every
block with a cost factor according to its refinement level (and thereby
to the number of integration steps), the fact that only blocks of a
certain level are integrated at a given time causes idle periods on
some nodes unless every node has the same number of blocks {\it for
every level}.

\begin{figure}
\centerline{\includegraphics[width=9cm]{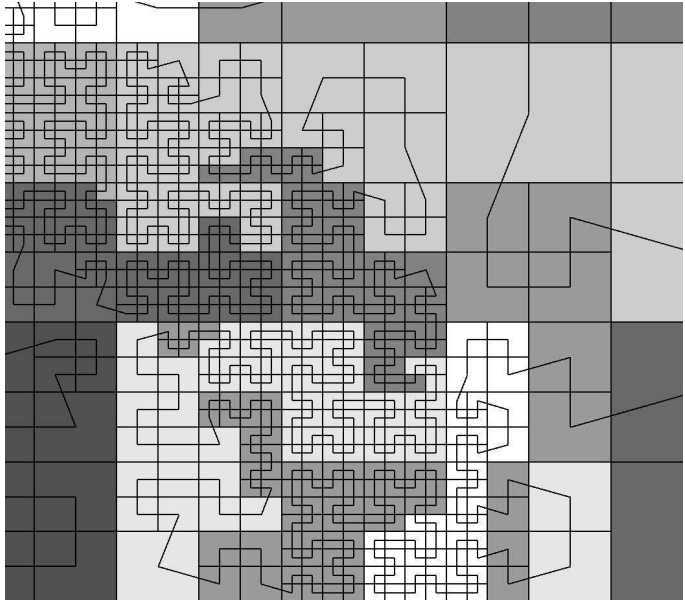}}
\caption
{ Close-up view of a number of blocks up to refinement level 6,
taken from an actual simulation run. The gray scale indicates the
computer node on which a block is located.  Each block consists of
16x16 cells in this case.  Distribution is based on the Hilbert
curve ordering on a per-level basis, which leads to the obvious
clustering of neighbouring blocks on each level.  }
\end{figure}

Figure 3 shows a close-up view of a local grid structure created
during a simulation run together with their ordering as induced by
the space filling curves.
The effect of cluster formation of blocks belonging to the same
refinement level (i.e. having the same size) is obvious, and remote
communication between computing nodes is then restricted to the lines
dividing these cluster.

In our implementation, load distribution is directly coupled to the
grid refinement:
After a regridding procedure, which might create or remove blocks on
the various nodes, the new load per node is calculated for every
level.
If the load imbalance exceeds a given threshold, blocks are migrated
between the nodes to arrive at the optimal load distribution.
Note that the distributed regridding already involves a collective
communication, because the refinement criterion applied to a block on
one node might flag a neighboring block, possible residing on a
different node, as described in section 3.
Therefore, grid refinement flags have to be exchanged between nodes
before the actual refinement takes place, which then is a local process.
Afterwards, the re-distribution is carried out, if necessary.

\subsection{POSIX-Multi-thread Parallelization}

On SMP machines, the use of multi-threading instead of distribution
into different processes seems a natural way for parallelization.
While OpenMP is an established standard here, \Rac uses explicit
thread programming through the standard POSIX interface in order to
preserve finer control over the workload distribution and the affinity
between CPUs and memory.
As mentioned earlier, a general observation is that the efficient
cache utilization  is crucial for high
computing performance on modern CPU architectures .
Therefore, the load distribution and balancing by means of Hilbert
curves has been carried over to the multi-thread parallelization in
that each POSIX thread (one per CPU) operates on a fixed subset of all
grid blocks according to the calculated distribution.
Moreover, each thread has its own copy of the entire grid (i.e. the
block tables and connectivity information) with the peculiarity that
only those blocks that are local to the thread are populated with
data.
Un-populated blocks act as proxies and contain pointers to the actual
data in the ``spheres'' of other threads if they're in the same address space.
All data allocation takes place from within the thread that later
operates on the block data in order to ensure that a good
CPU-to-memory affinity is achieved also on  architectures with non-uniform memory
access (NUMA).

With this strict data separation between the threads, not only the
numerical operations on the block data, but also the iterations
related to grid traversal and retrieval of connectivity information
from block objects operate on local copies for each thread.
Thus, the fundamental idea is to effectively create thread-private
data spaces and basically restrict all memory accesses that might
trigger cache-coherency controllers to enforce re-caching operations
to the inevitable exchange of boundary data between grid blocks.
As mentioned, this exchange is limited by means of the partitioning
according to space-filling curves.
A typical time step on the grid (or on one grid level, depending on
the integration mode) consists of first advancing all blocks (in which
the thread-allocated data is accessed ``privately'' by the working
thread), followed by data copy which involves overlapping memory access
and might cause cache conflicts.

\begin{figure}
\centerline{\includegraphics[width=9cm]{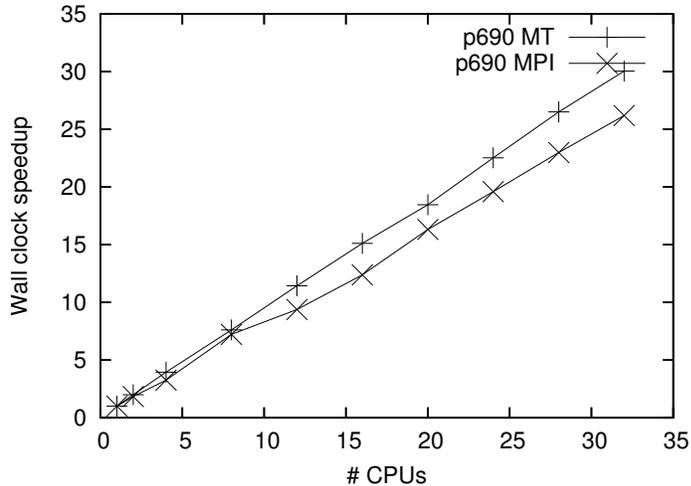}}
\caption
{ Wall-clock speed-up of the test application on IBM p690 with
multithread parallelization (MT) and single-thread MPI distributed
parallelization (MPI), compared to the serial run (at \# CPU=1). The
serial execution time is 2253 seconds, floating point rate is
$\approx$ 600 MFlips. 
Times exclude data initialization, I/O, regridding.
 }
\end{figure}

\begin{figure}
\centerline{\includegraphics[width=9cm]{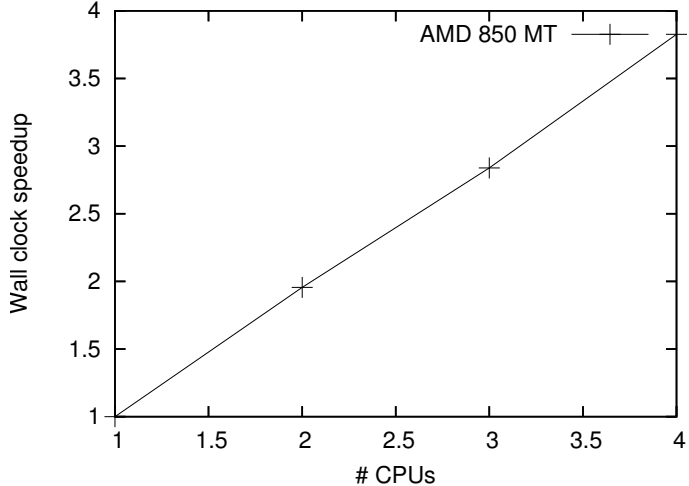}}
\caption
{ Wall-clock speed-up of the test application on AMD opteron 850 quad
with multithread parallelization (MT) compared to the serial run (\# CPUs = 1).
Here, the serial execution time (w/o initialization, I/O, regridding) is 176 seconds.
The single thread MPI measurements gave essentially the same results.
}
\end{figure}

We have tested this approach on the JUMP multiprocessor at
Forschungszentrum Jülich, which consists of IBM p690 machines, each with 32
Power4+ CPUs at 1.7 GHz clock rate running AIX, and on a local quad AMD
Opteron 850 machine with 2.4 GHz CPU clock running Linux with a 2.6.10 kernel.
Fig. 4 and 5 show the execution time of test runs with a typical
setup.
Here, the 3-dimensional Euler equations from gas dynamics used to
simulate gas blow-out from dwarf galaxies as a result of supernovae
explosions, are integrated for 200 time steps.
The numerical scheme is the Kurganov-Tadmor scheme with 3rd order
CWENO reconstruction and 2nd order Runge-Kutta integration on grid
blocks with $8^3$ cells and 2 layers of ghost cells, the number of
grid blocks was approximately 4800 on the p690 and 590 on the Opteron 850
quad setting.
The execution times given cover only the numerical integration itself,
without diagnostics, I/O, regridding or initialization of block data.
On the IBM p690 power4+, the measured sustained floating point rate
was around 600 MFlips per CPU, which is close to 9\% of the
theoretical peak performance and can be considered a reasonable value
for this kind of application.
The execution time for the communication alone (boundary data exchange
by copy, interpolation, averaging) amounts to about 10\% of the
total execution time in all cases.

In all cases, the scaling behavior is good to excellent, which, however,
was obtained only after further manual optimization:
In order to guarantee a tight task affinity between the executing
threads and the CPUs, and thereby take full advantage of the affinity
between threads and data as described above, we had to explicitly bind
the created threads to single CPUs in the machines.
In AIX, this was achieved with the ``bindprocessor'' system function,
on Linux with ``sched\_setaffinity.''
This manual tuning was essential for the displayed scaling behavior:
Without the enforced CPU affinity, the parallel performance degraded
by about $10--15\%$ on both platforms, with considerable fluctuations
between runs.
Binding the threads to CPUs, however, resulted in the well reproducible
scaling figures shown in  Figures 4 and 5.
The interpretation here is that (unwanted) spontaneous task migrations
of the computing threads between CPUs during the run leads to random
enhancements of the distances to the memory where the corresponding
data is allocated and to necessary re-cache operations.

\subsection{MPI Parallelization}

The parallelization over address spaces, necessary for network-based
clusters, is similar in principle to the method described above:
Here, every process hosts a subset of all grid blocks, but the data exchange
now involves message passing with MPI rather than direct memory copies.
An explicit serialization of message data has been implemented instead
of using the many specialized transfer functions and derived data
types of MPI.
The reason is again that many different kinds of messages are
exchanged with different parts of a ghost cell layer, and their flow
is hardly predictable, in particular for the exchange at refinement
boundaries.
It seems much more straight-forward to compose messages that contain
meta information about the message type, the target block and the
target region within the block (e.g.~which ghost cells) together with
the core data itself.

Two different approaches have been followed concerning the message size, number
and send time:
\begin{itemize}
\item[1.] Each data portion to be transferred (e.g. part of a ghost cell layer
of one specific block to one specific neighbor block) is composed into
a MPI message and posted asynchronously with MPI\_Isend as soon as the
sending block is updated.
The matching receive requests are posted on the receiving side {\it
before} the level update, and message completion and dispatch occurs
after the level update.
\item[2.] Small inter-block messages to the same target process are
collected into larger MPI messages of fixed sizes.  These are
transferred after a level update is finished and dispatched on the
basis of the embedded meta information. The need for continuation
messages is indicated by the sender through meta tags in the compound
message and triggers a second communication encounter between sending and
receiving process.
\end{itemize}
The first approach typically leads to many messages of small sizes
(for the test example computation described above, there are of the
order of $10^3$ messages with an average size of $\approx 1.5$ kB
{\sl per time step}), but the message exchange is initiated as soon as the
send data is ready, which in principle allows for concurrency between
computation and communication.

With the second method, messages are much larger (we chose 100 kB for
the compound message size), but they are exchanged only after a level
update is complete.
Message collection is performed into sliced buffers of fixed slice
size which grow on demand and are reset after the communication is completed.
At that point, the buffers are not de-allocated but get reused in the
next data exchange in order to avoid frequent allocation-deallocation
sequences.
A complication here is that the number of necessary slices within a
communication cycle is practically unpredictable a priori, as it
depends on the order in which the smaller inter-block data is
serialized due to padding effects.
One way to go here would be to compute the needed total message size
for every sender-receiver pair in advance and negotiate the resulting
MPI message size after each regridding operation.
This, however had to be done for every type of data exchange
separately (boundary copy, coarse-fine interpolation, fine-coarse
update, etc.)  and also for each boundary transfer object, as
different exchanges for different physical fields might be used
in more complex simulation settings.
For simplicity, we used a fixed MPI compound message size and took
into account that message padding leads to some communication overhead.

For a small number of processes, both approaches gave satisfactory
results, the speedup-curves for the second approach on the IBM
machines is given in Figure 4 for up to 32 processors.
As can be seen, the scaling is still reasonable up to 32 processors
(with a speedup factor of about 26), but slightly inferior to the pure
multi-threaded method.

Similar to the multi-thread case, a smaller test problem was run on the
Opteron 850 quad with the LAM MPI implementation.
This resulted in basically the same numbers as the multi-threaded case
(Fig. 5).
In all these cases, the MPI traffic was intra-machine and has been
communicated through shared memory inter-process communication by the
corresponding MPI transport layer.

\subsection{Hybrid Parallelization}

The two parallelization methods described so far have been combined
into a hybrid mode in the obvious way:
A number of processes are started, and within each, a configurable
number of threads is started up to work concurrently.
As before, each thread task within a process consists on a fixed
subset of grid blocks, with its own copy of the grid structure and
connectivity information.
Thereby, all threads work autonomously in large portions of the
computation, but are synchronized before data exchange operations.

Depending on the choice of the MPI data exchange mode, the
inter-process communication differs slightly:
\begin{itemize}
\item[1.]  When each inter-block data message is mapped into one MPI
message, all executing threads post their corresponding messages
immediately after a block is updated.  After a complete level update,
one thread finishes the MPI requests, and all threads dispatch the
received messages concurrently.  This method requires a thread-safe
MPI implementation and was tested with LAM MPI and LAMPI on Linux and
with the IBM implementation on AIX.
\item[2.] When collecting inter-block messages into compound MPI
messages, all threads serialize concurrently into the send buffers to
a specific peer process.
Here, the memory allocation and housekeeping within the compound
message is realized in a thread-safe way by standard
mutex-locking. Data serialization itself happens non-exclusively after
allocation.
After a level-update, only one main thread performs the MPI
send-receive operations and queues the received compound messages,
which are de-serialized and dispatched concurrently by the other
threads (or the main thread itself, if there is only one thread per process).
\end{itemize}

As with the pure MPI mode, both methods were tried on the p690
machines and the comparison resulted clearly in favor of the second
way, i.e. the use of compound messages for the MPI transfer.
This might be surprising in the first instance, as only one main
thread is in charge of the entire MPI communication (send/receive),
but it must be kept in mind that 1. the latency is fairly low (some
$\mu$s) and 2. that the serialization/de-serialization work is shared
between the threads.
Also, on the receiving side, de-serialization and MPI communication
can overlap from the second compound message on.

A further remark is due for the first method, i.e.~concurrent MPI traffic
from all threads:
First of all, the performance degraded seriously here, with an
increase in execution time of up to 20\% in some cases.
Modifying the method in a way that not only one thread performed the MPI
message completion, but all threads did concurrent waits, each on
those messages that target their ``thread-local''
blocks, gave even worse results and had to be discarded entirely.
This behavior indicates that the matching of many small  messages during
the wait stage from within different threads can cause severe
penalties, at least when done in the naive way it was implemented here
(where the matching of message destination and message type was
realized through the MPI message tag and no distinct MPI
communicators were used).
Also, even with thread-safe MPI implementations, the internals of the
communication layers and possible thread synchronization and
serialization is not visible to the user and depends strongly on the
implementation itself.

\begin{figure}
\centerline{\includegraphics[width=9cm]{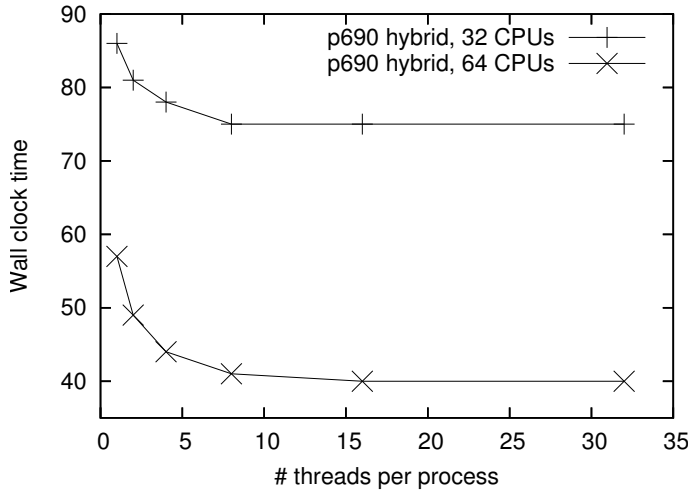}}
\caption
{ Wall-clock time for test application execution in
hybrid-parallelization on IBM p690 with 32 and 64 CPUs, respectively.
Displayed is the execution time with 1 to 32 threads per process.
The total number of threads is 32 and 64, respectively, grouped in
a variable number of processes.
}
\end{figure}

Speedup-curves for the hybrid mode obtained on the p690 are given in
Figure 6 for the previously discussed test run with 32 and 64 CPUs,
respectively.
Here, the total number of working threads equal the number of
available CPUs, but the grouping into processes ranges from the pure
MPI parallelization (1 thread per process, 32 resp.~64 processes) to
the maximum of 32 threads per process (i.e.~pure multi-thread mode in
the case of 32 CPUs or 2 processes with 32 threads spread over 2
machines in the case of 64 CPUs).
The MPI communication is based on compound messages of 100 kB size,
and the task affinity was again enforced through explicit CPU binding
as described in the multi-thread section.

It is obvious that there is some performance loss with increasing
numbers of processors, while the speed-up from 32 to 64 CPUs in the 32
threads case corresponds to a factor of 1.9 (76s to 40s).
Comparing the most favorable runs with the pure MPI runs, we get a
performance loss of 13\% in the 32 CPU case (76s to 86s), and a more
pronounced penalty of 42\% in the 64 CPU case (40s to 57s).
These results indicate that the hybrid mode in fact can lead to a
gain compared to the pure MPI parallelization once the multi-threaded
execution is optimized by enforcing the tight affinity between tasks,
CPUs and memory.

\section{Conclusions}

We have described the development of a new framework for mesh-adaptive
computations of hyperbolic conservative systems.
The project has been inspired by the attempt to create a modular basis
for AMR computations which can easily be extended and specialized, and 
which is suited for parallel computing environments.

Through the use of object-oriented methods, the complexity associated
with AMR could be implemented successfully with limited efforts.
During the iterative development cycle, we noticed that the principles
of code re-use and abstraction saved us considerable efforts when
changing parts of the program or correcting errors.

A regular subgridding method, now widely used \cite{Flash},
\cite{Powell}, \cite{DAGH}, has been proven to be a
straight-forward but well working approach for the mesh refinement,
and the recent development of new numerical schemes for hyperbolic
systems, which offer an attractive combination of simplicity,
robustness and accuracy, provided a further major ingredient in the
philosophy of a flexible framework.

As for parallelization, we have developed a hybrid concept of
multi-threading and inter-process communication through MPI.
The benchmarks on IBM p690 machines with 32 CPUs show that the hybrid
concept in fact results in performance gain over a pure MPI
parallelization, which, however requires a careful optimization of the
multi-threaded implementation.

A key issue here is the efficient use of the CPU cache, which in the
first place can be naturally obtained in AMR by the use of small grid
block sizes that fit well into the cache.
In addition, each thread in the current implementation creates its own
effective data subspace consisting of a fixed subset of grid blocks
and the block connectivity information, all allocated in the thread
itself in order to achieve small CPU-memory distances in NUMA
architectures.
Block assignment to threads is based on the same space filling curve
algorithm that determines the distribution among processes, and which
thereby tends to minimize not only inter-process communication but
also inter-thread memory accesses with potential cache conflicts.
To finally achieve the desired gain from multi-threading, the affinity
between tasks and CPUs must be enforced manually by binding the
working threads to individual CPUs.

For the MPI part of the communication, it turned out that the creation
of fewer messages of moderate size (1MB and below) by collecting the
small inter-block messages which are addressed to the same target
processor is favorable compared to mapping the typically small
messages ($\approx$ one to few kB) between blocks directly to MPI
messages, despite the fact that all MPI traffic is channeled through
one thread in the message collection method, 
Here, the (de-) serialization of compound MPI messages can occur
concurrently by many threads.
These results indicate that the concurrent access to the MPI layer for
the completion of many small-sized messages, even with multi-thread
abilities, should be used carefully with respect to the overall
performance.

In the end, the hybrid concept proved to work satisfactory and
resulted in floating point performances in the range of 7--10\% of the
theoretical peak performance on 64 processors for the described
application.
Naturally, there is still some room for further improvement, for
example in connection with automatic estimates for the size of
MPI compound messages.
For practical use, further development of high-level interfaces for
the control of task and memory affinity on high performance computers
would be helpful, as the method of explicit CPU binding that was chosen
here has the potential to conflict with the job dispatcher and load distribution 
algorithms in larger settings.
One interesting initiative for IBM's platform is the VSRAC interface project
(www.redbooks.ibm.com/redpapers/pdfs/redp3932.pdf), that might
be extended in the near future to allow a thread-level control in
addition to its current process-level control.

With respect to the performance and scaling figures given here, it
should be mentioned once more that the ratio of computation to
communication costs with the tailored numerical schemes discussed here
is very favorable in that the communication accounts for no more than
10\% of the execution time.
Other schemes with more global data dependencies and more volume data
communication, like for example elliptical solvers, might be more
sensible to the communication overhead and therefore may require
an increased effort to optimize the parallelization.
These further investigations are planned for an adaptive multigrid
solver which is currently under development.

\ack 
We would like to thank the members of the Zentralinstitut f\"ur
Angewandte Mathematik at Forschungszentrum J\"ulich for their
assistance and advice on issues of the parallel implementation.
Access to the JUMP multiprocessor computer at Forschungszentrum Jülich
was made available through project HBO18.
This work benefitted from support through INTAS contract 00-292 and
Sonderforschungsbereich 591 of Deutsche Forschungsgemeinschaft.
We thank an anonymous referee for helpful comments and suggestions on
the original manuscript.

\end{document}